\documentstyle[12pt,epsfig]{article}
\begin{document}
\setlength{\parskip}{0.45cm}
\setlength{\baselineskip}{0.75cm}
\begin{titlepage}
\begin{flushright}
DO-TH 96/11 \\ RAL-TR-96-034  \\ June 1996
\end{flushright}
\vspace{0.6cm}
\begin{center}
\Large
\hbox to\textwidth{\hss
{\bf Spin-Dependent Parton Distributions } \hss}

\vspace{0.1cm}
\hbox to\textwidth{\hss
{\bf of the Longitudinally Polarized Photon } \hss}

\vspace{0.1cm}
\hbox to\textwidth{\hss
{\bf Beyond the Leading Order} \hss}

\vspace{1.2cm}
\large
M.\ Stratmann\\
\vspace{0.5cm}
\normalsize
Universit\"{a}t Dortmund, Institut f\"{u}r Physik, \\
\vspace{0.1cm}
D-44221 Dortmund, Germany \\
\vspace{1.2cm}
\large
W. Vogelsang \\
\vspace{0.5cm}
\normalsize
Rutherford Appleton Laboratory \\
\vspace{0.1cm}
Chilton, Didcot, Oxon OX11 0QX, England \\
\vspace{1.6cm}
{\bf Abstract}
\vspace{-0.3cm}
\end{center}
A next-to-leading order (NLO) QCD analysis of the spin-dependent
parton distributions $\Delta f^{\gamma}(x,Q^2)$ of the
longitudinally polarized photon and of its structure function
$g_1^{\gamma}(x,Q^2)$ is performed within the framework of the
radiative parton model. The important issues of a suitably
chosen factorization scheme and related boundary conditions
are discussed in detail. The typical effects of the NLO corrections
are quantitatively studied for two very different conceivable scenarios
for the NLO polarized parton distributions $\Delta f^{\gamma}(x,Q^2)$.
\end{titlepage}
%
%MAIN PART
%
%INTRODUCTION
%
\noindent
More accurate measurements of the nucleon's spin asymmetry
$A_1^N(x,Q^2)$ $\simeq$  $g_1^N(x,Q^2)/$ $F_1^N(x,Q^2)$ in polarized
deep-inelastic scattering (DIS) \cite{ref1}, covering also a wider
range in $(x,Q^2)$ and providing results for different targets
($N=n,d$) as compared
to early measurements of $A_1^p(x,Q^2)$ \cite{ref2}, have
considerably improved our
knowledge about the nucleon's spin structure in the past few years
and also renewed
the theoretical interest in this field. This is also due to
the possibility to perform now a complete and consistent
QCD analysis of polarized DIS in NLO, since the required
spin-dependent two-loop splitting functions 
$\Delta P_{ij}^{(1)}$ have been calculated recently \cite{ref3,ref4}.
A first such NLO analysis in the $\overline{\rm{MS}}$ scheme
has been presented in \cite{ref5}
based on the phenomenologically successful concept of the
radiative parton model, i.e., the generation of parton
distributions from a valence-like structure at some
low-resolution scale $\mu$, which had previously led, e.g.,
to the prediction \cite{ref6} of the small-$x$ rise of the unpolarized
proton structure function $F_2^p(x,Q^2)$ as observed at
HERA \cite{ref7}. Subsequent NLO studies \cite{ref8} have imposed different
boundary conditions and/or factorization schemes.

The knowledge of the two-loop splitting functions
$\Delta P_{ij}^{(1)}$ \cite{ref3,ref4}
also offers the opportunity to perform a similar NLO QCD analysis of
the spin-dependent parton content $\Delta f^{\gamma}$ of the
longitudinally (more precisely, circularly) polarized {\em{photon}} 
because the required two-loop
photon-to-parton splitting functions $\Delta k_q^{(1)} \equiv 
\Delta P_{q\gamma}^{(1)}$ and $\Delta k_g^{(1)} \equiv \Delta 
P_{g\gamma}^{(1)}$ can be easily obtained from $\Delta P_{qg}^{(1)}$
and $\Delta P_{gg}^{(1)}$, respectively. Although such a study seems to
be somewhat premature in view of the lack of {\em{any}} experimental
information on $\Delta f^{\gamma}$ up to now, interesting theoretical
questions arise when going beyond the leading order. Apart from getting
a feeling for the typical size of the NLO corrections it is moreover
important to analyse the necessity (and feasibility) to introduce a suitable
factorization scheme which overcomes expected problems with perturbative 
instabilities arising in the $\overline{\rm{MS}}$ scheme in particular for 
large values of $x$. Such instabilities were found in the 
unpolarized case where they were eliminated \cite{ref9} by absorbing the
the 'direct-photon' contribution to $F_2^{\gamma}$ into the NLO photonic 
quark distributions (${\rm{DIS}}_{\gamma}$ scheme). In this paper
we will show that a similar procedure is also recommendable in the polarized 
case, where it works equally well. 

Furthermore it is no longer inconceivable to longitudinally polarize
also the proton beam at
HERA \cite{ref11}. At such high energies the polarized electron acts
dominantly as a source of almost real (Weizs\"acker-Williams) photons,
thus measurements of double spin asymmetries in, e.g., the photoproduction
of large-$p_T$ jets can in principle reveal information on the parton
content of the polarized photon in addition to that of the proton
\cite{ref12} through the presence of resolved-photon processes. In the
corresponding situation with unpolarized beams this has been already
extensively studied experimentally \cite{ref7}. Future polarized
linear $e^+e^-$ colliders could serve to provide additional
complementary information on $\Delta f^{\gamma}$ \cite{ref18} by measuring the
spin-dependent photon structure function $g_1^{\gamma}(x,Q^2)$ or
spin asymmetries in resolved two-photon reactions.

In the remainder of the paper we present all necessary ingredients for the
two-loop evolution of the spin-dependent parton distributions of the photon
and for the calculation of its structure function $g_1^{\gamma}$ in NLO, 
analysing also the aforementioned theoretical questions.  
We will work within the framework of the radiative parton
model since the corresponding analysis for the unpolarized photon
\cite{ref10} has again been phenomenologically very successful \cite{ref7}.
We will present two 'extreme' sets of polarized NLO distributions
$\Delta f^{\gamma}(x,Q^2)$ following closely a previous LO analysis
\cite{ref13,ref18}. 

%
%TECHNICAL SECTION
%
Similarly to the purely hadronic case it is convenient to decompose 
the spin-dependent parton distributions $\Delta f^{\gamma}(x,Q^2)$
$(f=u,\,d,\,s,\,g)$ of the longitudinally polarized photon into
flavor non-singlet (NS) quark combinations $\Delta q_{NS}^{\gamma}$ and
the singlet (S) part
$\Delta \vec{q}^{\,\gamma}_{S} \equiv {\Delta \Sigma^{\gamma}
\choose \Delta g^{\gamma}} $, where
$\Delta \Sigma^{\gamma} \equiv \sum_f (\Delta f^{\gamma}+
\Delta \bar{f}^{\gamma} )$ with $f$ running over all relevant active
quark flavors and $\Delta g^{\gamma}$ denotes the polarized photonic 
gluon distribution. The so defined combinations $\Delta q_i^{\gamma}(x,Q^2)$
($i$=NS, S) satisfy the well-known
{\em inhomogeneous} evolution equations schematically
given by\footnote{
We follow closely the notation adopted in the unpolarized
case as presented in refs.\cite{ref14} and \cite{ref9}.}
\begin{equation}
\label{gl1}
\frac{d \Delta q_i^{\gamma}(x,Q^2)}{d \ln Q^2} = \Delta k_i(x,Q^2)+
\left( \Delta P_i \ast \Delta q_i^{\gamma} \right) (x,Q^2) \;\; ,
\end{equation}
where the symbol $\ast$ denotes the usual convolution in
Bjorken-$x$ space which reduces, in Mellin-$n$ space, to a
simple product $\Delta P_i^n \Delta q_i^{\gamma ,n}$ with the
$n$th moment of a function $h(x,Q^2)$ being defined as
\begin{equation}
\label{gl2}
h^n(Q^2) \equiv \int_0^1 x^{n-1} h(x,Q^2)\, dx \;\;\;.
\end{equation}
The polarized photon-to-parton and parton-to-parton splitting functions,
$\Delta k_i(x,Q^2)$ and $\Delta P_i(x,Q^2)$, respectively,
in eq.(\ref{gl1}) receive the following 1-loop (LO) and 2-loop (NLO) 
contributions ($i$=NS, S):
\begin{eqnarray}
\label{gl3}
\nonumber
\Delta k_i(x,Q^2) &=& \frac{\alpha}{2 \pi} \Delta k_i^{(0)}(x)+
\frac{\alpha \alpha_s(Q^2)}{(2\pi)^2} \Delta k_i^{(1)}(x) \\
\Delta P_i(x,Q^2) &=& \frac{\alpha_s(Q^2)}{2 \pi}
\Delta P_i^{(0)}(x) + \left(\frac{\alpha_s(Q^2)}{2 \pi}
\right)^2 \Delta P_i^{(1)}(x)  \;\; ,
\end{eqnarray}
where $\alpha\simeq 1/137$ and the NLO running strong coupling
is given by
\begin{equation}
\label{gl4}
\frac{\alpha_s(Q^2)}{4\pi} \simeq \frac{1}{\beta_0 \ln Q^2/
\Lambda_{\overline{\rm{MS}}}^2} - \frac{\beta_1}{\beta_0^3}
\frac{\ln \ln Q^2/\Lambda_{\overline{\rm{MS}}}^2}
{\left(\ln Q^2/\Lambda_{\overline{\rm{MS}}}^2\right)^2}
\end{equation}
with $\beta_0=11-2 N_f/3$, $\beta_1=102-38 N_f/3$, and $N_f$ being the
number of active flavors.
In the singlet (S) case eq.(\ref{gl1})
becomes, of course, a coupled $2\times 2$ matrix equation where
\begin{equation}
\renewcommand{\arraystretch}{1.5}
\label{gl5}
\Delta \hat{P}_{S}^{(j)} \equiv \left[ \begin{array}{cc}
\displaystyle \Delta P_{qq}^{(j)} &
\displaystyle \Delta P_{qg}^{(j)} \\
\displaystyle \Delta P_{gq}^{(j)} &
\displaystyle \Delta P_{gg}^{(j)}
\end{array} \right]\;\;\; ,\;\;\;
\Delta \vec{k}_{S}^{(j)} = \left( \begin{array}{c}
\displaystyle \Delta k_q^{(j)} \\
\displaystyle \Delta k_g^{(j)} \end{array}\right)
\end{equation}
\noindent
in eq.(\ref{gl3}) with $j=0,1$. The hadronic polarized splitting functions
$\Delta P_{ff'}^{(j)}$ can be found in \cite{ref3,ref4} and apart
from obvious NS and S charge factors,
$\langle e^4\rangle-\langle e^2\rangle^2$
and $\langle e^2\rangle$, respectively, where
$\langle e^k\rangle \equiv N_f^{-1} \sum_{i=1}^{N_f} e_{q_i}^k$,
the spin-dependent photon-to-parton splitting function
$\Delta k_q^{(0)}$ can be obtained from $\Delta P_{qg}^{(0)}$
by multiplying it with $N_f N_C/T_F$ where $N_C=3$ and $T_F=N_f/2$;
similarly the NLO quantities $\Delta k_q^{(1)}$ and
$\Delta k_g^{(1)}$ correspond to the $C_F T_F$ terms
of $\Delta P_{qg}^{(1)}$ and $\Delta P_{gg}^{(1)}$,
respectively, multiplied by $N_f N_C/T_F$ :\footnote{
Note that $\Delta k_g^{(0)}=0$ due to the missing 
photon-gluon coupling in lowest order. Furthermore, there
is a subtlety in deriving $\Delta k_g^{(1)}$ from 
$\Delta P_{gg}^{(1)}$ because the latter splitting
function is a diagonal quantity and hence contains
$\delta (1-x)$ terms originating from gluon self-energy
contributions which have to be omitted in 
$\Delta k_g^{(1)}$ \cite{ref9,ref15}.}
\begin{eqnarray}
\label{gl6}
\nonumber
\Delta k_{NS}^{(j)}(x) &=& N_f N_C (\langle e^4\rangle -\langle e^2\rangle^2)
\Delta \kappa^{(j)}(x)\;,\;\;\;
\Delta k_q^{(j)}(x)=N_f N_C \langle e^2\rangle \Delta \kappa^{(j)}(x)\\
\nonumber
\Delta \kappa^{(0)}(x) &=& 2 \left[ x^2-(1-x)^2\right] \\
\nonumber
\Delta \kappa^{(1)}(x) &=& C_F \, \Bigg[ -9\ln x
+8(1-x)\ln (1-x)+27x-22\\
\nonumber
&+& \frac{1}{2} \Bigg\{\ln^2 x+2\ln^2 (1-x)-4\ln x \ln (1-x)-\frac{2}{3}\pi^2
\Bigg\} \Delta \kappa^{(0)} (x) \Bigg]\\
\nonumber
\Delta k_g^{(0)}(x) &=& 0\\
\Delta k_g^{(1)}(x) &=& N_f N_C \langle e^2\rangle C_F \left[
-2(1+x) \ln^2 x +2(x-5)\ln x-10(1-x)\right]\;\;\; ,
\end{eqnarray}
where $C_F=4/3$.

The evolution equations (\ref{gl1}) 
are most conveniently solved directly in
Mellin-$n$ space where the solutions can be given analytically and one
can easily keep track of the contributions stemming from different powers
of $\alpha_s$ in order to avoid terms beyond the order considered.
Taking, according to eq.(\ref{gl2}), the $n$th moment of 
eq.(\ref{gl1}) the various
convolutions simply factorize and the required moments of the photonic
inhomogeneous LO and NLO $\Delta k$ terms in 
eqs.(\ref{gl1},\ref{gl3},\ref{gl5},\ref{gl6}) are given by
\begin{eqnarray}
\label{gl7}
\nonumber
\Delta k_{NS}^{(j)n} &=& N_f N_C (\langle e^4\rangle -\langle e^2\rangle^2)
\Delta \kappa^{(j)n}\;,\;\;\;
\Delta k_q^{(j)n} = N_f N_C \langle e^2\rangle \Delta \kappa^{(j)n}\\
\nonumber
\Delta \kappa^{(0)n} &=& 2 \,\frac{n-1}{n(n+1)}\\
\nonumber
\Delta \kappa^{(1)n} &=& C_F \Delta \kappa^{(0)n} \Bigg[
S_1(n)^2-S_2(n)-\frac{2}{n} S_1 (n)
+ \frac{5n^4+10n^3-n+2}{2 n^2(n+1)^2} \Bigg]\\
\nonumber
\Delta k_g^{(0)n} &=& 0\\
\Delta k_g^{(1)n} &=& N_f N_C \langle e^2\rangle C_F \, \Bigg[-2\,
\frac{(n-1) (n+2) (n^2-n-1) }{n^3(n+1)^3}\Bigg]\;\;\;.
\end{eqnarray}
with $S_k(n)\equiv \sum_{j=1}^{n} j^{-k}$. The Mellin moments 
$\Delta P_{ff'}^{(j)n}$ of the 1- and 2-loop hadronic splitting functions can 
be found in \cite{ref5}\footnote{Note that the $n$th moments of the hadronic
splitting functions $\Delta P_{ff'}^{(j)n}$ $(j=0,\,1)$ and the anomalous
dimensions $\Delta \gamma_{ff'}^{(j)n}$ as given in \cite{ref3,ref5} are
related through $\Delta P_{ff'}^{(0)n}=-\frac{1}{4}
\Delta \gamma_{ff'}^{(0)n}$ and $\Delta P_{ff'}^{(1)n}= -\frac{1}{8}
\Delta \gamma_{ff'}^{(1)n}$.} in a form appropriate for a straightforward 
analytic continuation in $n$ (also given in \cite{ref5}) which is required 
for a numerical Mellin inversion back into $x$-space. 
The solution of eq.(\ref{gl1}) can be decomposed into a 'pointlike' 
(inhomogeneous\footnote{By definition, we choose the pointlike part to 
satisfy $\Delta q_{i,PL}^{\gamma ,n}(\mu^2)=0$ 
($i=$NS,S) at the input scale $\mu$.}) and a 'hadronic' (homogeneous) part, 
i.e.,
\begin{equation}
\label{gl8}
\Delta q_i^{\gamma ,n}(Q^2) = \Delta q^{\gamma ,n}_{i,PL}(Q^2) +
\Delta q^{\gamma ,n}_{i,had}(Q^2)
\end{equation}
($i=$ NS, S) and can be found in \cite{ref9}
(with the obvious replacements of all
unpolarized quantities like, e.g., $k_i^{(1)n}$, by the corresponding
polarized ones, e.g., $\Delta k_i^{(1)n}$); they need not be repeated
here. Having solved the evolution equations (\ref{gl1}) for
$\Delta q_{NS}^{\gamma ,n}(Q^2)$, $\Delta \Sigma^{\gamma ,n}(Q^2)$,
and $\Delta g^{\gamma ,n}(Q^2)$ one finally obtains the desired
photonic parton distributions $\Delta f^{\gamma ,n}(Q^2)$ $(f=u,\,
d,\,s,\,g)$ by a straightforward flavor decomposition.

In moment-$n$ space the NLO expression for the spin-dependent
photon structure function $g_1^{\gamma}$ is given by
\begin{eqnarray}
\label{gl9}
\nonumber
g_1^{\gamma ,n}(Q^2) &=& \frac{1}{2} \sum_{f=u,d,s} e_f^2\;
\Bigg\{ \Delta f^{\gamma ,n}
(Q^2) + \Delta\bar{f}^{\gamma ,n}(Q^2)  \\
\nonumber
&+& \frac{\alpha_s (Q^2)}{2\pi} \left[ \Delta C_q^n \left(
\Delta f^{\gamma ,n}(Q^2) + \Delta\bar{f}^{\gamma ,n}(Q^2) \right) +
\frac{1}{N_f} \Delta C_g^n \Delta g^{\gamma ,n}(Q^2)\right] \Bigg\}  \\
&+& \frac{1}{2} N_f N_C \langle e^4\rangle \frac{\alpha}{2\pi} 
\Delta C_{\gamma}^n
\end{eqnarray}
with the usual hadronic spin-dependent Wilson coefficients $\Delta C_q^n$ and
$\Delta C_g^n$ which in the conventional $\overline{\rm{MS}}$ scheme  
can be found, e.g., in ref.\cite{ref5}. The photonic coefficient
$\Delta C_{\gamma}^n$ can be easily derived from $\Delta C_g^n$ and is
in the $\overline{\rm{MS}}$ scheme given by:
\begin{equation}
\label{gl10}
\Delta C_{\gamma}^n = \frac{1}{T_F} \Delta C_g^n = 2 \left[ -
\frac{n-1}{n(n+1)} \left( S_1(n)+1\right) -\frac{1}{n^2} +
\frac{2}{n(n+1)} \right]
\end{equation}                                     
corresponding to the $x$-space expression
\begin{equation}
\label{gl11}
\Delta C_{\gamma}(x) = 2 \left[(2x-1) \left(\ln \frac{1-x}{x} -1\right)
+2(1-x)\right]\;\;\;.
\end{equation}
We note that the LO expression for $g_1^{\gamma}$ is
entailed in the above formula (\ref{gl9}) by simply dropping
all NLO terms, i.e., all $\Delta C_i^n$ $(i=q,\; g,\; \gamma)$.
For what follows it is convenient to introduce the decomposition 
of $g_1^{\gamma ,n}(Q^2)$ into a pointlike and a hadronic part,
analogously to eq.(\ref{gl8}): 
\begin{equation}
\label{g1decomp}
g_1^{\gamma ,n}(Q^2) \equiv g_{1,PL}^{\gamma ,n}(Q^2) + 
g_{1,had}^{\gamma ,n}(Q^2)  \; ,
\end{equation}
where $g_{1,PL}^{\gamma ,n}(Q^2)$ is obtained from eq.(\ref{gl9})
by taking only $\Delta f^{\gamma ,n}(Q^2)=\Delta f^{\gamma ,n}_{PL}(Q^2)$ 
with $\Delta f^{\gamma ,n}_{PL}(Q^2)$ as defined in (\ref{gl8}).
Conversely, for $g_{1,had}^{\gamma ,n}(Q^2)$ one uses the 
$\Delta f^{\gamma ,n}_{had}(Q^2)$ of (\ref{gl8}), and one obviously has to
omit the $\Delta C_{\gamma}^n$ term in (\ref{gl9}) in this case.

The desired $x$-space expressions for $\Delta f^{\gamma}(x,Q^2)$ and
$g_1^{\gamma}(x,Q^2)$ can be easily obtained from the above given $n$-space
expressions $\Delta f^{\gamma ,n}(Q^2)$ and $g_1^{\gamma ,n}(Q^2)$,
respectively, by performing a standard numerical Mellin inversion.

%%%%%%%%%%%%%%%%%%%%%%%%%%%%%%%%
% SCHEME AND BOUNDARY CONDITIONS
%%%%%%%%%%%%%%%%%%%%%%%%%%%%%%%%
The solutions for $\Delta f^{\gamma ,n}(Q^2)$ ($\Delta f^{\gamma}(x,Q^2)$)
depend on the up to now unspecified hadronic input distributions at the
input scale $Q^2=\mu^2$, i.e., on the boundary conditions for the hadronic
pieces $\Delta f_{had}^{\gamma,n}$ in (\ref{gl8}) which one would 
intuitively relate to some model inspired by vector meson dominance (VMD).
On the other hand, beyond LO both the 'pointlike' as well as the 
'hadronic' pieces in (\ref{gl8}) depend on the factorization scheme chosen,
and it is a priori not clear in which type of factorization schemes
it actually makes sense to impose a pure VMD hadronic input. Indeed, in the
unpolarized case it was observed that \cite{ref9} the $\ln (1-x)$ term in 
the photonic coefficient function $C_{2,\gamma}(x)$ for $F_2^{\gamma}$,
which becomes negative and divergent for $x\rightarrow 1$, 
drives the pointlike part of $F_2^{\gamma}(x,Q^2)$ in the $\overline{\rm{MS}}$
scheme to large negative values as $x\rightarrow 1$, leading to a strong
difference between the LO and the NLO results for $F_{2,PL}^{\gamma}$ in the
large-$x$ region. As illustrated in Fig.1, a very similar thing happens 
in the polarized case: Here it is the $\ln (1-x)$ term in the polarized 
photonic coefficient function $\Delta C_{\gamma}(x)$ (see eq.(\ref{gl11}))
for $g_1^{\gamma}$ that causes large negative values of the pointlike part 
of $g_1^{\gamma}(x,Q^2)$ in the $\overline{\rm{MS}}$ scheme as 
$x\rightarrow 1$, strongly differing from the corresponding LO result 
also shown in Fig.1. Clearly, the addition of a VMD-inspired hadronic part 
$\Delta f^{\gamma ,n}_{had}(Q^2)$ cannot be sufficient to cure this observed 
instability of $g_{1,PL}^{\gamma}$ in the large-$x$ region since any VMD 
input vanishes as $x\rightarrow 1$. Instead,
as in the unpolarized case, an appropriately adjusted ('fine tuned')
non-VMD hadronic NLO input would be required in the $\overline{\rm{MS}}$ 
scheme, substantially differing from the LO one, as the only means of 
avoiding unwanted and physically not acceptable perturbative instabilities for 
physical quantities like $g_1^{\gamma}(x,Q^2)$.

In the unpolarized case the so-called $\rm{DIS}_{\gamma}$ scheme \cite{ref9}
was introduced to avoid such 'inconsistencies' by absorbing the photonic
Wilson coefficient for $F_2^{\gamma}$ into the photonic quark distributions.
Analogously, one expects that a similar procedure for the coefficient
$\Delta C_{\gamma}$ for $g_1^{\gamma}$ cures the problem observed 
for $g_{1,PL}^{\gamma}$ in the $\overline{\rm{MS}}$ scheme. This 
redefinition of the polarized photonic quark distributions implies, of course,
also a transformation of the NLO photon-to-parton splitting functions 
$\Delta k_i^{(1)}$ due to the requirement that the physical quantity 
$g_1^{\gamma}$ has to be scheme independent. In the polarized case the 
transformation to the $\rm{DIS}_{\gamma}$ scheme reads 
\begin{equation} \label{ctraf}
\Delta C_{\gamma}^n\rightarrow \Delta C_{\gamma}^n +
\delta \Delta C_{\gamma}^n \; \; , 
\end{equation}
where $\delta \Delta C_{\gamma}^n=-
\Delta C_{\gamma}^n$. This implies for the $\Delta k_i^{(1)n}$ ($i=$NS,S)
in eq.(\ref{gl7}) that $\Delta k_i^{(1)n}\rightarrow \Delta k_i^{(1)n} +
\delta \Delta k_i^{(1)n}$ with \cite{ref9}
\begin{eqnarray}
\label{gl12}
\nonumber
\delta \Delta k_{NS}^{(1)n} &=& - N_f N_C (\langle e^4\rangle -
\langle e^2\rangle^2) \Delta P_{qq}^{(0)n} \Delta C_{\gamma}^n \\
\delta \Delta \vec{k}_S^{(1)n} &=&
\left( \begin{array}{c}
\displaystyle \delta \Delta k_q^{(1)n} \\
\displaystyle \delta \Delta k_g^{(1)n} \end{array}\right)
= - N_f  N_C \langle e^2\rangle 
\left(\begin{array}{c}
\displaystyle \Delta P_{qq}^{(0)n} \Delta C_{\gamma}^n \\
\displaystyle \Delta P_{gq}^{(0)n} \Delta C_{\gamma}^n
\end{array}
\right) \;\;\;.
\end{eqnarray}
It should be emphasized that all hadronic quantities, in particular 
$\Delta C_q^n$ and $\Delta C_g^n$, are unaffected by this kind of
scheme transformation. We remark that if one chooses to solve the evolution 
equations for the $\rm{DIS}_{\gamma}$ polarized photonic parton distributions 
$\Delta f^{\gamma}(x,Q^2)$ directly in $x$-space by a (cumbersome) numerical 
iterative procedure the Mellin inverse of $\delta \Delta k_i^{(1)n}$ in 
eq.(\ref{gl12}) is explicitly needed. Using standard integrals \cite{ref17}
and \cite{ref9}
\begin{equation}
\label{gl14}
\int_0^1 \;dx\;x^{n-1} \left[\ln^2(1-x)-\ln x \ln(1-x) -{\rm{Li}}_2(x)
\right]=\frac{1}{n}\left[S_1(n)\right]^2
\end{equation}
one obtains for
$\Delta \kappa_q^n\equiv \Delta P_{qq}^{(0)n} \Delta C_{\gamma}^n$
\begin{eqnarray}
\label{gl15}
\nonumber
\Delta \kappa_q(x)\!\!\! &=& \!\!\!C_F  
\Bigg[ - 7 + 4x + (-5+8x)
\ln x + (15-16 x)\ln (1-x) \\
\!\!\!&+&\!\!\! 
(2x-1) \left[4\ln^2(1-x)-4\ln(1-x)\ln x+\ln^2 x + 2 {\rm{Li}}_2(x)
-\pi^2 \right]
\!\!\Bigg]  \; \; ,
\end{eqnarray}
and the inverse of $\Delta\kappa_g^n\equiv \Delta P_{gq}^{(0)n}
\Delta C_{\gamma}^n$ reads
\begin{eqnarray}
\label{gl16}
\Delta \kappa_g(x) &=& 2 C_F \Bigg[ -12 (1-x) +(-7+x)\ln x -
(1+x) \ln^2 x + (1+x) \frac{\pi^2}{3} \nonumber \\
&+& 5(1-x) \ln(1-x) -2 (1+x) {\rm{Li}}_2(x) \Bigg]\;\;\;.
\end{eqnarray}
Inspecting eqs.(\ref{gl6}),(\ref{gl12}),(\ref{gl15}),(\ref{gl16}) 
one finds that the transformation to the $\rm{DIS}_{\gamma}$ scheme,
besides curing the instabilities at $x\rightarrow 1$, also 
eliminates all terms $\sim \ln^2 x$ from the polarized NLO photon-to-parton 
splitting functions $\Delta k_i^{(1)} (x)$ ($i=$NS,S), i.e., removes 
the $\overline{\rm{MS}}$ terms leading for $x\rightarrow 0$
(for corresponding observations in the unpolarized case see 
\cite{lund,grvfrag}).      

The result for $g_{1,PL}^{\gamma}$ after the transformation   
to the $\rm{DIS}_{\gamma}$ scheme is also shown in Fig.1. The similarity
between the NLO ($\rm{DIS}_{\gamma}$) and the LO curves strongly suggests 
that it is indeed recommendable also in the polarized case to work in the 
$\rm{DIS}_{\gamma}$ scheme. We note that it turns out, however, that the 
resulting $g_{1,PL}^{\gamma}$ slightly exceeds the pointlike part of the 
unpolarized photon structure function $F_{1,PL}^{\gamma}$ in the vicinity
of $x\sim 0.6$, thus making a violation of the fundamental positivity
constraint $|g_{1}^{\gamma}|\leq F_{1}^{\gamma}$ imminent there. The underlying
reason for this feature is not a defect of the $\rm{DIS}_{\gamma}$ scheme
as such, but resides in the fact that in the unpolarized case the 
$\rm{DIS}_{\gamma}$ scheme was formulated \cite{ref9} in terms of 
(the only measured structure function)
$F_2^{\gamma}$, and {\em not} $F_1^{\gamma}$. The difference between the 
unpolarized photonic coefficient functions $C_1^{\gamma}$ and
$C_2^{\gamma}$ (for $F_1^{\gamma}$ and $F_2^{\gamma}$, respectively)
decreases $F_{1,PL}^{\gamma}$ with respect to $F_{2,PL}^{\gamma}/2x$,
which explains the above effect. The problem could be straightforwardly 
resolved by repeating the analysis of \cite{ref9,ref10} in a modified 
$\rm{DIS}_{\gamma}$ scheme for which one would choose to absorb 
$C_1^{\gamma}$ rather than $C_2^{\gamma}$ into the unpolarized NLO 
photonic quark densities. This is clearly beyond the scope of this
paper. We mention in this context that in the unpolarized case also an 
alternative factorization scheme was suggested \cite{aur} for which only the 
'process independent' part of the photonic Wilson 
coefficient for $F_2^{\gamma}$ 
is absorbed into the photonic quark distributions. This scheme partly 
shares the properties of the $\rm{DIS}_{\gamma}$ scheme to warrant a 
reasonable behaviour 
of $F_{2,PL}^{\gamma}$ in the large-$x$ region. In the polarized 
case it is easy to see that the ansatz of \cite{aur} amounts to 
transforming $\Delta C_{\gamma}^n$ via eq.(\ref{ctraf}) by
\begin{equation}
\delta \Delta C_{\gamma}^n = -2 \left[ - \frac{n-1}{n(n+1)} S_1(n) +
\frac{2}{n(n+1)^2} \right] \; \; ,
\end{equation}
with corresponding changes of $\Delta k_i^{(1)n}$. For completeness we
include the result for $g_{1,PL}^{\gamma}$ in this factorization scheme 
in Fig.1. It turns out that the above mentioned slight violation of 
positivity does not occur if both the polarized and unpolarized NLO quark 
densities are defined in this scheme. On the other hand, it becomes 
obvious that a significant dissimilarity between the 
LO and NLO results remains, which would demand compensation by sizeably 
different LO/NLO hadronic inputs\footnote{Similar features were observed for
this scheme in the unpolarized case \cite{lund}.}. We therefore do not
pursue this scheme any further, but will henceforth adopt the 
$\rm{DIS}_{\gamma}$ scheme as introduced above when studying the polarized 
photon structure beyond the leading order.

For convenience, we provide the relation of the NLO $\rm{DIS}_{\gamma}$ and
$\overline{\rm{MS}}$ photonic parton distributions since it is to be expected 
that future calculations of NLO corrections to polarized cross sections will 
be carried out within the $\overline{\rm{MS}}$ scheme. The 
$\Delta f^{\gamma}$ in the $\overline{\rm{MS}}$ scheme can be obtained 
by the transformation
\begin{equation}
\label{gl22}
\Delta f^{\gamma}_{\overline{\rm{MS}}}(x,Q^2) =
\Delta f^{\gamma}_{\rm{DIS}_{\gamma}}(x,Q^2)+\delta \Delta f^{\gamma}(x,Q^2)
\end{equation}
with
\begin{equation}
\label{gl23}
\delta \Delta q^{\gamma}(x,Q^2)=
\delta \Delta \bar{q}^{\gamma}(x,Q^2) = -N_C e_q^2 \frac{\alpha}{4 \pi}
\Delta C_{\gamma}(x)\;,\;\;\;\delta \Delta g^{\gamma}(x,Q^2)=0  \; ,
\end{equation}
where $\Delta C_{\gamma}(x)$ is given in eq.(\ref{gl11})\footnote{
Alternatively, of course, one can work directly with the 
$\rm{DIS}_{\gamma}$ distributions by applying an appropriate transformation
\cite{ref9} to NLO sub-cross sections calculated in the $\overline{\rm{MS}}$ 
scheme for processes involving polarized real photons.}.
                                 
To finish this technical part of the paper, we briefly discuss the 
so-called NLO 'asymptotic' solution for the spin-dependent parton 
distributions of the photon, which is obtained by dropping
all terms in the full (pointlike) solution which decrease with increasing 
values of $Q^2$. In this way all dependence on the input scale and the 
boundary conditions is eliminated, and one ends up with the 
unique QCD prediction
(see \cite{asy,ref14,ref9,vphd} for a discussion of the asymptotic   
solution in the unpolarized case)
\begin{equation}
\label{asympt}
\Delta \vec{q}^{\gamma ,n}_{PL}(Q^2) = \frac{4\pi}{\alpha_s (Q^2)} 
\Delta \vec{a}^n + \Delta \vec{b}^n  \; ,
\end{equation}
where 
\begin{eqnarray}
\label{asy1}
\Delta \vec{a}^n &=& \frac{1}{1-(2/\beta_0) \Delta \hat{P}^{(0)n}} 
\frac{\alpha}{2\pi\beta_0} \Delta \vec{k}^{(0)n}  \nonumber \; , \\
\Delta \vec{b}^n &=& -\frac{1}{\Delta \hat{P}^{(0)n}} \left[ 
2 \left( \Delta \hat{P}^{(1)n} - \frac{\beta_1}{2\beta_0} \Delta 
\hat{P}^{(0)n} \right) \Delta \vec{a}^n + \frac{\alpha}{2\pi} 
\left( \Delta \vec{k}^{(1)n} - \frac{\beta_1}{2 \beta_0} \Delta 
\vec{k}^{(0)n} \right) \right]\!.
\end{eqnarray}
The above equations have been written for the singlet case; extension
to the non-singlet sector is trivial. The polarized LO asymptotic solution, 
which was already studied in \cite{xu}, is entailed in the expressions by 
dropping all NLO terms, i.e., keeping the $\Delta \vec{a}$ term only.
The NLO asymptotic parton densities in 
(\ref{asympt}) are obviously again subject to the factorization 
convention adopted. For instance, one could choose to work in the
$\overline{\rm{MS}}$ scheme for which the $\Delta \vec{k}^{(1)n}$ is
as given in (\ref{gl7}), or again in the $\rm{DIS}_{\gamma}$ where 
the transformation (\ref{gl12}) is to be taken into account.
However, unlike the non-asymptotic pointlike solution $g_{1,PL}^{\gamma}$, 
the asymptotic prediction for $g_1^{\gamma}$, to be obtained from
eqs.(\ref{asympt}),(\ref{gl9}), is readily seen to be scheme-independent
up to terms of ${\cal{O}}(\alpha_s)$,
as it has to be. It is also displayed in Fig.\ 1 for
$Q^2=20\,{\rm{GeV}}^2$. The practical utility of the
asymptotic solution is very limited, since it only applies at very large $Q^2$
and $x$. Furthermore, the determinants of the denominators 
$1-(2/\beta_0) \Delta \hat{P}^{(0)n}$ and $\Delta \hat{P}^{(0)n}$    
in (\ref{asy1}) can vanish, causing completely unphysical poles of the 
asymptotic solution which are {\em not} present in the full solution where
subleading ('non-asymptotic') terms regulate such pole terms.
These obvious defects of the asymptotic solution are well-known from 
the unpolarized case \cite{ref14,ref9} and need not be discussed again in 
detail. We only mention that the position of the poles in Mellin-$n$
space can differ from the unpolarized case due to the in general different
splitting functions involved: For $N_f=3$ flavors the determinant of 
$1-(2/\beta_0) \Delta \hat{P}^{(0)n}$ vanishes for $n=0.2903$ 
in the non-singlet case\footnote{Needless to say that in this case 
$\Delta \hat{P}^{(0)n} \rightarrow \Delta P_{qq}^{(0)n}= P_{qq}^{(0)n}$.}
and for $n=0.3673$ and $n=1$ in the singlet
case. It turns out, however, that the pole at $n=1$ is cancelled twice
by terms in the numerator, such that the LO asymptotic solution 
has a vanishing first moment. The situation becomes worse at NLO, 
where the poles arising from $1/\Delta \hat{P}^{(0)n}$ have to be taken into
account. Since the first moment of $\Delta P_{qq}^{(0)}$ vanishes,
the non-singlet solution has a potential pole at $n=1$. As in LO, it
is cancelled by terms from the numerator, but this time the result remains 
{\em finite} at $n=1$, such that the NS part of the asymptotic 
solution has a non-vanishing first moment beyond LO. In the singlet sector, 
the determinant of $\Delta \hat{P}^{(0)n}$ develops a zero at 
$n=1.5723$ ($N_f=3$). This implies that the singlet asymptotic solution
will rise as $\approx x^{-1.57}$ as $x\rightarrow 0$, i.e., the polarized NLO
asymptotic photonic parton densities, as well as the asymptotic result
for $g_1^{\gamma}$, will not be integrable anymore. This clearly underlines
that the asymptotic solution can in general not be regarded as a reliable 
or realistic estimate for the polarized photon structure.

%%%%%%%%%%%%%%%%
% PHENO. SECTION
%%%%%%%%%%%%%%%%
To study more quantitatively the influence of the QCD corrections we extend a
previous LO analysis of the polarized photon structure within the radiative
parton model \cite{ref13,ref18} to NLO in the $\rm{DIS}_{\gamma}$ factorization 
scheme as described above. As pointed out above, the main advantage of the 
$\rm{DIS}_{\gamma}$ scheme is \cite{ref10} that an optimal perturbative
stability is achieved for the pointlike part of the photonic structure 
functions $F_2^{\gamma}$ and $g_1^{\gamma}$, implying that no additional 
'fine-tuned' input is required in NLO. One thus expects that the hadronic 
inputs in LO and NLO will differ by just the small amounts known from 
similar analyses of nucleon structure functions (see, e.g., \cite{ref6}, and
\cite{ref5} for the polarized case), and that beyond the LO
the $\rm{DIS}_{\gamma}$ scheme is the most likely scheme in which
a {\em pure VMD} hadronic input can be successfully implemented.
In fact, such a result was found in the unpolarized case in \cite{ref10}, 
where, starting the evolution from a low input scale, it was shown
that rather similar hadronic VMD-inputs were sufficient in LO and NLO 
to describe existing data for $F_2^{\gamma}$ at larger $Q^2$ accurately.
Since nothing is known experimentally about the parton structure 
of vector mesons, the parton densities of the neutral pion as determined in
a previous study \cite{grvp} were used instead which are expected not 
to be too dissimilar from those of, e.g., the $\rho$. 
Unfortunately, such a procedure is obviously impossible for determining 
the VMD input distributions $\Delta f^{\gamma}(x,\mu^2)$ for the 
{\em polarized} photon. Therefore, to obtain a realistic estimate for the 
theoretical uncertainties in the polarized photon structure functions 
coming from the unknown hadronic input, two very different scenarios 
were considered in \cite{ref13,ref18}: For the first ('maximal scenario') 
the input was characterized by
\begin{equation}
\label{gl18} 
\Delta f^{\gamma}_{had}(x,\mu^2) = f_{had}^{\gamma}(x,\mu^2)
\end{equation}
whereas the other extreme input ('minimal scenario') was defined by
\begin{equation}
\label{gl19}
\Delta f^{\gamma}_{had}(x,\mu^2) = 0
\end{equation}
with $\mu^2=\mu^2_{LO}=0.25\,{\rm{GeV}}^2$ and the unpolarized LO
distributions $f_{had}^{\gamma}(x,\mu^2)\equiv$ \linebreak $f_{had,LO}^{\gamma}
(x,\mu^2_{LO})$ taken from \cite{ref10}. We will closely follow
this approach and thus in NLO ($\rm{DIS}_{\gamma}$) take
$\mu^2=\mu^2_{NLO}=0.3\,{\rm{GeV}}^2$ and the unpolarized NLO
densities $f_{had}^{\gamma}(x,\mu^2)=f_{had,NLO}^{\gamma}(x,\mu^2_{NLO})$ 
from \cite{ref10} in eqs.\ (\ref{gl18}) and (\ref{gl19}) to define the two 
extreme scenarios. We mention at this point that a sum rule expressing the 
vanishing of the first moment of the polarized photon structure function 
$g_1^{\gamma}$ was derived from current conservation in \cite{sr}, which 
we could use to further restrict the range of allowed VMD inputs.
The sum rule can be realized in LO and NLO ($\overline{\rm{MS}}$ or 
$\rm{DIS}_{\gamma}$) by demanding 
\begin{equation}
\label{quark}
\Delta q_{had}^{\gamma,n=1}(\mu^2)=0 \:\:\: ,
\end{equation}
i.e., a vanishing first moment of the photonic quark densities at the
input scale. Inspecting the relevant LO and NLO evolution kernels 
and coefficient functions for $n=1$, in particular the expressions for
the $\Delta k_i^{(1)n}$ in (\ref{gl7}) and (\ref{gl12}), one finds that the 
sum rule $g_1^{\gamma,n=1} (Q^2)=0$ is then maintained for {\em all} $Q^2$ even 
beyond the LO\footnote{As mentioned above, this is no longer true for
the NLO asymptotic solution.}. Of the two extreme hadronic inputs introduced 
above only the 'minimal' one (eq.(\ref{gl19})) satisfies (\ref{quark}).
On the other hand, we are interested only in the region of, say,
$x>0.01$ here, such that for the 'maximal' scenario (\ref{gl18}) the 
current conservation constraints at the input scale could well be 
implemented by contributions from smaller $x$ which do not affect, of course,
the evolutions at larger $x$. In addition to this,
the first moment of the polarized photonic gluon distribution remains 
completely unconstrained by current conservation considerations. 
Rather than artificially
enforcing the vanishing of the first moment of the $\Delta q_{had}^{\gamma}
(x,\mu^2)$ in the 'maximal' scenario (see \cite{ref13}), we therefore 
stick to the two extreme scenarios as introduced above.

This fully specifies our polarized photonic NLO
($\rm{DIS}_{\gamma}$) distributions $\Delta f^{\gamma}(x,Q^2)$ for all
$Q^2\ge \mu^2$. The values for the QCD scale parameter 
$\Lambda_{\overline{\rm{MS}}}$ in NLO, appearing in eq.(\ref{gl4}) and 
used in the evolution equations, are also taken from \cite{ref10}, i.e.,
\begin{equation}
\label{gl20}
\Lambda^{(3,4,5)}_{NLO} = 248,\; 200,\; 131\;\;{\rm{MeV}} \; .
\end{equation}
We adopt all threshold conventions as in \cite{ref10} and our LO analysis
\cite{ref13,ref18}. 

In Fig.2 we compare our LO \cite{ref13,ref18} and NLO ($\rm{DIS}_{\gamma}$)  
distributions $x \Delta u^{\gamma}/\alpha$, $x \Delta g^{\gamma}/\alpha$ 
for the two extreme scenarios at $Q^2=10\,{\rm{GeV}}^2$. As can be seen,
the NLO distributions in the $\rm{DIS}_{\gamma}$ scheme are very similar
to the LO ones. Fig.3 shows the photonic structure function 
$x g_1^{\gamma}/\alpha$ in LO and NLO as calculated according to 
eq.(\ref{gl9}). Very satisfactory perturbative stability is found.
The result is presented for $N_f=3$ flavors, i.e.,
we have not included the charm contribution to $g_1^{\gamma}$ which could
be calculated via the polarized 'direct' fusion subprocess 
$\gamma^* \gamma \rightarrow c\bar{c}$ and the (small) 'resolved' process
$\gamma^* g \rightarrow c\bar{c}$ in which the polarized photonic
gluon distribution takes part \cite{ref18}. 
The charm contribution is immaterial for our
more illustrative purposes.

%%%%%%%%%
% SUMMARY
%%%%%%%%%
To summarize, we have provided all ingredients for a NLO analysis 
of the spin-dependent parton distributions of the photon and of its
polarized structure function $g_1^{\gamma}$. We have shown that 
$g_1^{\gamma}$ suffers from the same perturbative instability problems
as the corresponding unpolarized structure function $F_2^{\gamma}$ in the
$\overline{\rm{MS}}$ scheme which hampers a straightforward NLO analysis.
As we have demonstrated, it is therefore recommendable to work in a 
'polarized version' of the $\rm{DIS}_{\gamma}$ factorization scheme 
originally introduced in the unpolarized case in order to circumvent such 
problems. We have finally presented two extreme sets of polarized photonic
NLO parton distributions $\Delta f^{\gamma}(x,Q^2)$.

%%%%%%%%%%%%%%%%%
%Acknowledgements
%%%%%%%%%%%%%%%%%
We are grateful to M.\ Gl\"{u}ck and A.\ Vogt for many helpful discussions.
The work of M.S.\ has been supported in part by the
'Bundesministerium f\"{u}r Bildung, Wissenschaft, Forschung und
Technologie', Bonn.
%

%
%\newpage
%%%%%%%%%%%%%%%%%%%%%%%%%%%%%
\section*{Figure Captions}
\begin{description}
\item[Fig.1] The 'pointlike' part of $x g_1^{\gamma}/\alpha$ (see 
eq.(\ref{g1decomp})) in LO and NLO for the $\overline{\rm{MS}}$ and the
$\rm{DIS}_{\gamma}$ factorization schemes. Also shown is the result obtained
when extending the factorization scheme of \cite{aur} to the polarized case
('AFG', see text). The toy input scale $\mu=1$ GeV, the QCD scale 
parameter $\Lambda=200$ MeV and $N_f=3$ flavors have been used.
For illustration the NLO asymptotic solution as obtained from
eqs.(\ref{asympt}), (\ref{asy1}) is included in the lower part for
$Q^2=20\,{\rm{GeV}}^2$.
\item[Fig.2] Predictions for the NLO ($\rm{DIS}_{\gamma}$) polarized 
photonic parton densities for the 'maximal' and 'minimal' inputs 
of eqs.(\ref{gl18}) and (\ref{gl19}), respectively. For comparison we
also show the corresponding LO results of \cite{ref13,ref18}.    
\item[Fig.3] NLO predictions for the spin-dependent photon structure function
$g_1^{\gamma}$ for the 'maximal' and 'minimal' inputs of eqs.(\ref{gl18}) 
and (\ref{gl19}), respectively. The results shown correspond to $N_f=3$ 
flavors. For comparison we also present the respective LO predictions 
of \cite{ref13,ref18}.       
\end{description}
\newpage
\pagestyle{empty}

\vspace*{-2.1cm}
\hspace*{-0.7cm}
\epsfig{file=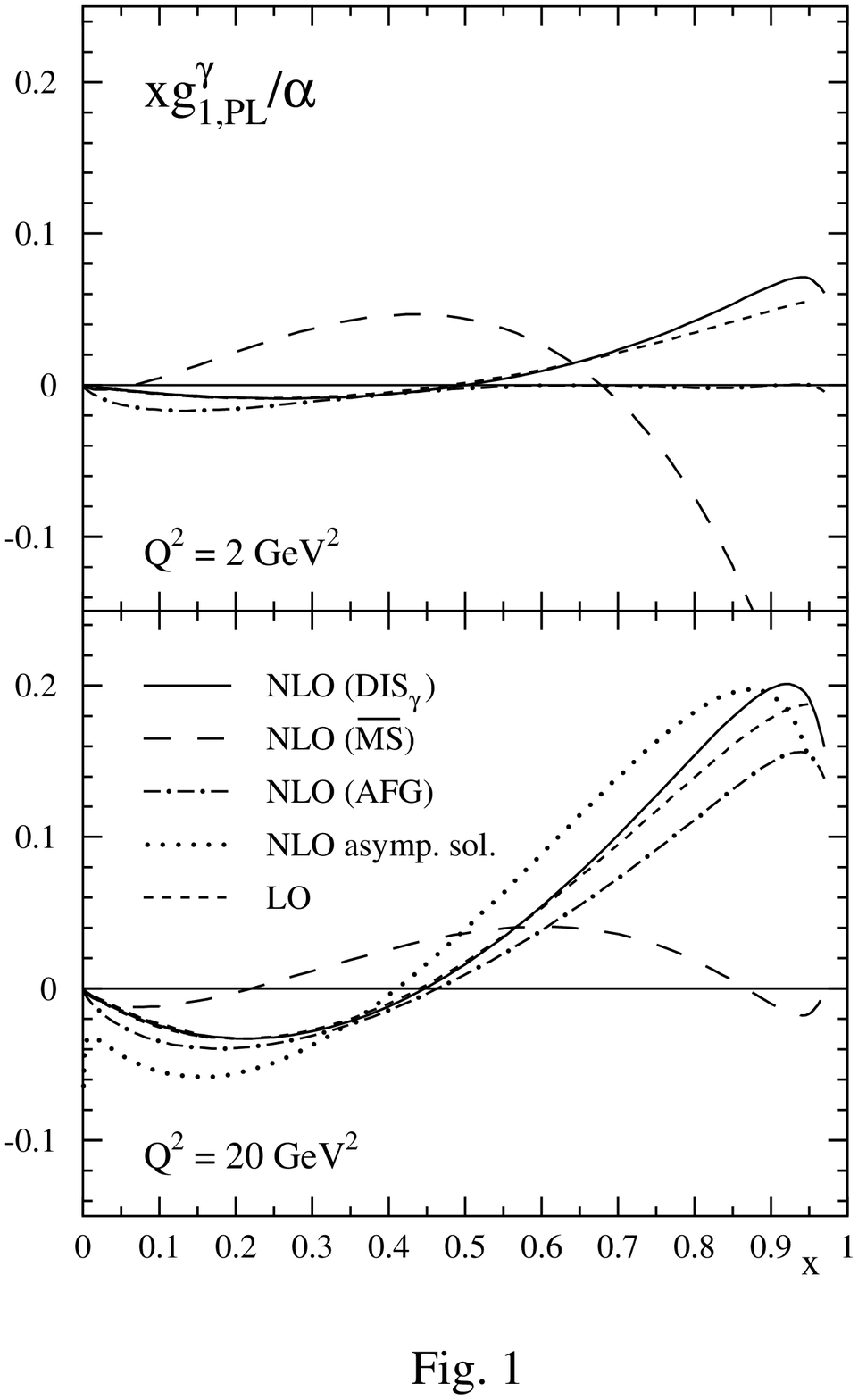}

\newpage
\vspace*{-1cm}
\hspace*{-1.0cm}
\epsfig{file=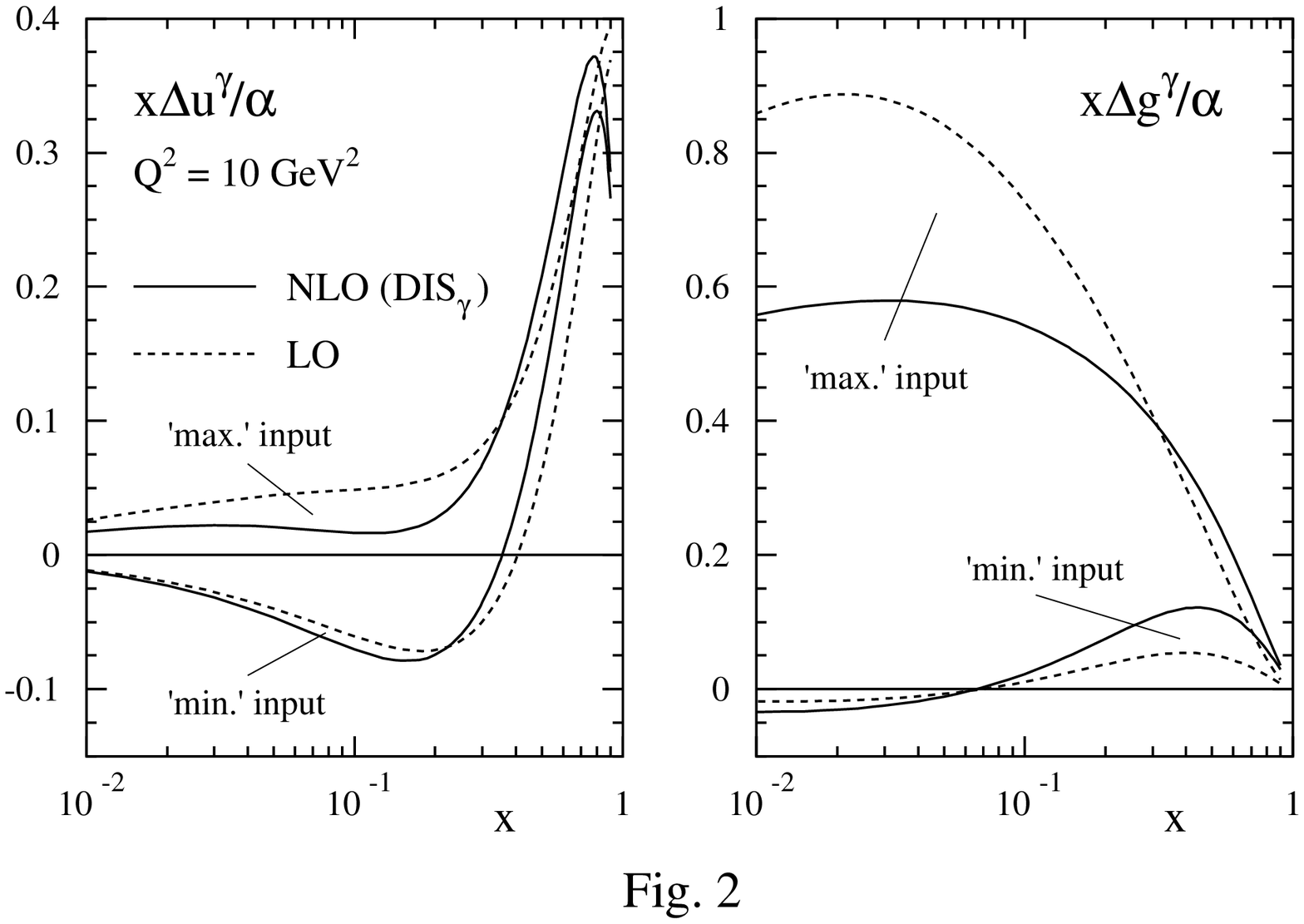,angle=90}

\newpage
\vspace*{1cm}
\hspace*{-1.3cm}
\epsfig{file=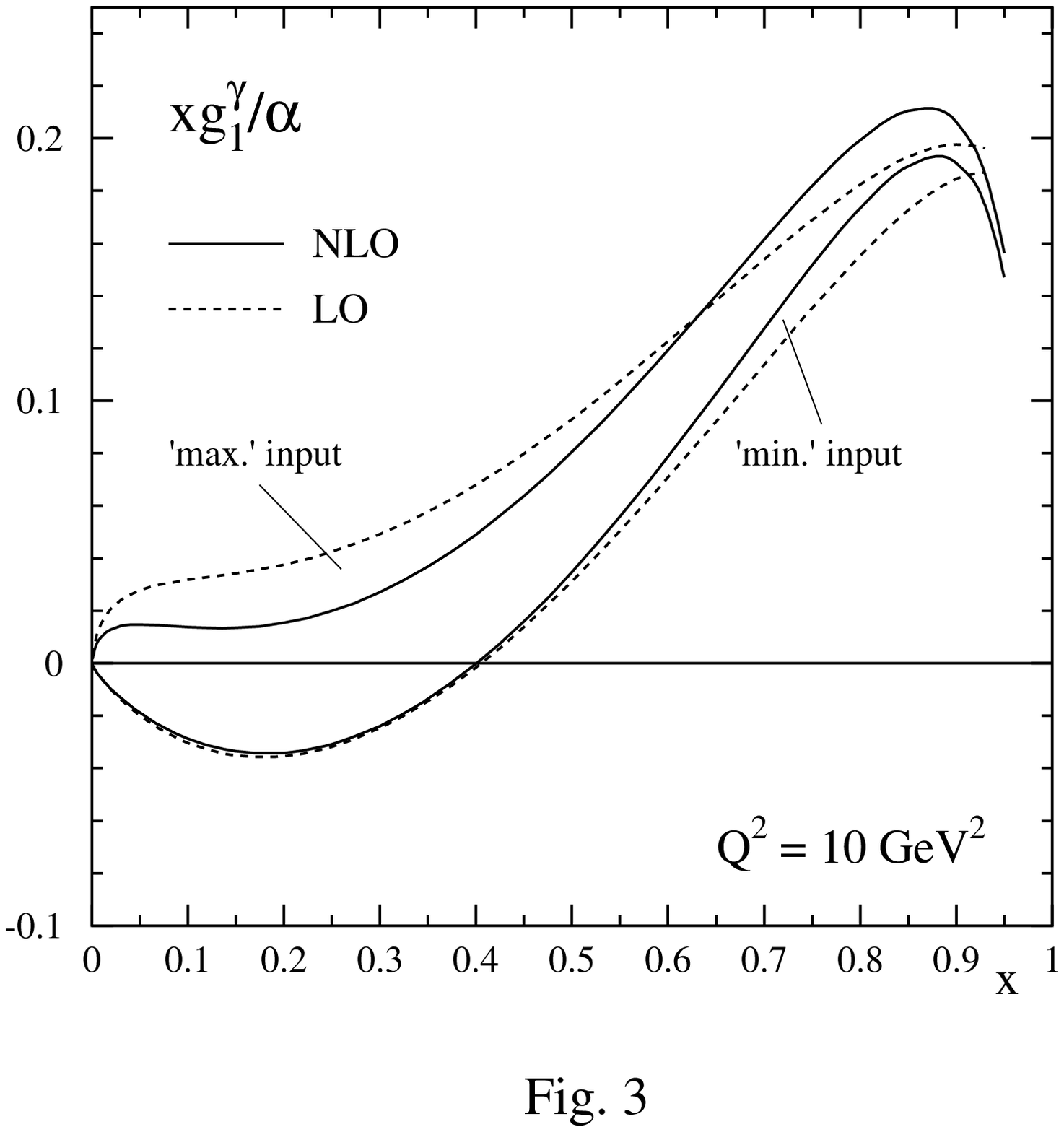}
\end{document}